# Do two parties represent the US?
## Clustering analysis of US public ideology survey

Louisa Lee[1] and Siyu Zhang[2, 3]
Advised by: Vicky Chuqiao Yang[1]

[1] Department of Engineering Sciences and Applied Mathematics, Northwestern University
[2] Department of Mathematics, Northwestern University
[3] Department of Electrical Engineering and Computer Science, Northwestern University

## Abstract
Recent surveys have shown that an increasing portion of the US public believes the two major US parties adequately represent the US public opinion and think additional parties are needed [1]. However, there are high barriers for third parties in political elections. In this paper, we aim to address two questions: "How well do the two major US parties represent the public's ideology?" and "Does a more-than-two-party system better represent the ideology of the public?". To address these questions, we utilize the American National Election Studies Time series dataset [2]. We perform unsupervised clustering with Gaussian Mixture Model method on this dataset. When clustered into two clusters, we find a large centrist cluster and a small right-wing cluster. The Democratic Party's position (estimated using the mean position of the individuals self-identified with the parties) is similar to that of the centrist cluster, and the Republican Party's position is between the two clusters. We investigate if more than two parties represent the population better by comparing the Akaike Information Criteria for clustering results of the various number of clusters. We find that additional clusters give a better representation of the data, even after penalizing for the additional parameters. This suggests a multiparty system represents of the ideology of the public better.

## Introduction
From the government shutdown in 2013 to the 14-hour filibuster on gun control in 2016, it is evident that cooperation between the political parties of the United States is steadily decreasing. The growing fissure between the Democratic and Republican ideologies and stances on issues is causing gridlock in the U.S. government. Frustrated by the lack of change, Americans are finding it difficult to support either political party. In a recent survey, 53% of likely US voters think it is fair to say that neither party in Congress is the party of the American people. Even 52% of those who identified as Republicans and 44% of those who identified as Democrats agree that neither major political party is the party that represents the American people [3]. This lack of confidence in the U.S. government bring into question the role of political parties and the principles of a democratic system to represent individuals and be a government of the people, by the people and for the people [4].

In this study, we aim to address two questions: "How well do the two major US parties represent the public opinion?" and "Does more than two parties better represent the ideology landscape?". In the past,

Contact information:
Louisa Lee: louisalee2019@u.northwestern.edu
Siyu Zhang: siyuzhang2018@u.northwestern.edu
Vicky Chuqiao Yang: vcy@u.northwestern.edu

there have been attempts to determine the political typologies of the American public. In 1986, Fleishman used cluster analysis to determine that there are six clusters of political attitudes that best represent the opinions of Americans. However, the dataset used was small (only 483 individuals) and outdated [5]. The Pew Research Center also conducted a study in 2014 that concluded with eight groups of political attitudes. However, the technique used to find these clusters is ambiguous and not detailed in their results [6].

Here, we use the Gaussian Mixture Model to cluster individuals' positions on political issues in the American National Election Studies dataset. The large sample size of data combined with the compatible method of clustering provides a representative model of America's public opinion space.

We conducted two types of clustering analysis. First, we fix the number of clusters to two, motivated by the current two-party system in the US, which is reinforced by the single-member district plurality election process and other barriers for additional parties to entry. Second, we allow for additional clusters, and use the Akaike Information Criteria (AIC) to determine whether more than two clusters represent the public opinion better.

## Data and Methods

We use the American National Election Studies [2] 1948-2012 time series dataset. The dataset includes survey results from more than 20,000 individuals from 1948 to 2012. The variables include self-reported positions on various political issues as well as party affiliation.

Cluster analysis was performed on the dataset using the Gaussian Mixture Model, a machine-learning technique that compares the likelihood of a set of data being generated by a number of clusters. This model uses the Expectation-Maximization algorithm to find the parameters of a Gaussian distribution (party center and a covariance matrix) that maximizes its likelihood. When comparing how the different number of clusters represent the data, we calculate and compare the AIC, which penalized additional parameters, to select the best model. We use MATLAB R2016a to perform the analysis.

Mathematically, the Gaussian Mixture Model is expressed as follows,

$$p(x|\mu_1 \ldots \mu_m, \Sigma_1 \ldots \Sigma_m) = \sum_{i=1}^{m} \alpha_i N_i(x|\mu_i, \Sigma_i),$$

where $m$ is the number of clusters, $\alpha_i$ is the weight of each cluster, $x$ is the observations in data, $\mu_i$ is the cluster center position of cluster $i$, and $\Sigma_i$ is the covariance matrix. $N_i(x|\mu_i, \Sigma_i)$ is the multivariate Gaussian distribution defined as,

$$N_i(x|\mu_i, \Sigma_i) = \frac{1}{(2\pi)^{D/2} |\Sigma_i|^{1/2}} \exp\{-\tfrac{1}{2}(x-\mu_i)^T \Sigma^{-1}(x-\mu_i)\}$$

where $D$ is the number of dimensions.

The cluster analysis was conducted on three subsets of data. The first includes opinions of individuals from 2012 on two political issues to simplify the model into two dimensions. The second contains opinions of individuals from 2012 on eleven political issues to represent the multidimensionality of opinion spaces. The third subset includes opinions of individuals from 1990 to 2012 on the three political issues to observe trends over time. Individuals' positions on political issues are reported on a seven-point scale. See Appendix A for detailed explanations of the political issues and corresponding survey questions. Only individuals with a complete set of data for each subset were included. The sample size for 2012 is 5914 individuals and the total sample size from years 1990 to 2012 is 20,502 individuals. For the seven-point scale questions, there were also options for "Don't Know" and "NA". These answers were assigned to the value of four, the median of the seven-point scale because they represent a group of individuals who are neutral about a specific political issue.

In the clustering analysis, we assumed that the multiple political issue dimensions are unrelated. This lead to the covariance matrix, $\Sigma_i$, (with the variance on the diagonals), are diagonals only. This treatment not only reduces the number of parameters that the model must calculate, but also prevents diagonal clusters from forming.

# Results

## Two Cluster Analysis

**Year 2012, Two Issues**

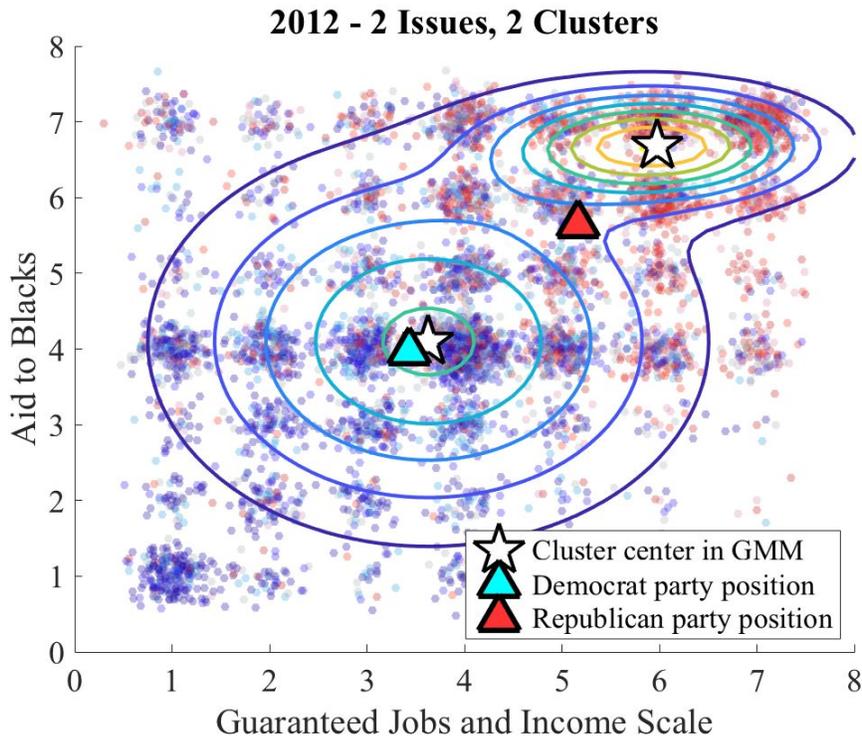

**Figure 1: Year 2012 – Two Issues, Two Clusters.** Ideology landscape over two issues (x and y-axis). Each marker represents an individual. The individuals self-identified as Republican, Democrat, and independent are marked in red, blue, and grey respectively. The darker the color, the stronger the party identification. The stars mark the two cluster means found by the Gaussian Mixture Model. The two triangles represent the political party means, estimated by averaging of the ideological positions of those that self-identified with each party.

|  | **Cluster 1** | **Cluster2** |
| --- | --- | --- |
| Cluster center position | (3.6 , 4.1) | (6.0 , 6.7) |
| Variance | (2.5 , 2.2) | (1.0 , 0.2) |
| Contains portion of data | 73% | 27% |
| Democrat | 49% | 16% |
| Independent | 36% | 39% |
| Republican | 15% | 45% |

**Table 1: Year 2012 – Two issues, Two Clusters.** This table includes resulting center position and variance for the clusters found using the Gaussian Mixture Model, the portion of data each cluster contains, and the composition of each cluster in terms of self-identified party affiliation.

The results from the two cluster analysis on two political issues in 2012 are shown in Figure 1. The x and y-axis represent two different political issues, and each dot on the graph is an individual. The political issue on the x-axis is Guaranteed Jobs and Income Scale while the political issue on the y-axis is Aid to Blacks Scale. See Appendix A for details on each political issue. Because the data are discrete, a small amount of noise was added to the coordinates to visualize the data. Each datum is also color coded to represent what party the individual self-identified with. Party affiliation was self-reported on a 7 point scale; Strong Democrat, Democrat, Independent-Democrat, Independent, Independent-Republican, Republican, and Strong Republican. Republicans are represented by shades of red, Independents by grey, and Democrats by blue. We calculate the party means of the Democratic and Republican parties by averaging the positions of the individuals who identify with the parties. Table 1 displays the center positions and variances of the clusters found, as well as the size and party composition of the clusters.

The clustering result suggests that the public's opinion on these two issues, when clustered into two groups, is best described by a large centrist cluster (containing 72% of the individuals) and a small right-wing cluster (containing 27% of the individuals). The Democratic Party's position is near the center of the centrist cluster, while the Republican Party's position is between the two clusters. The current landscape of two similarly sized parties may not be the best representation of the public opinion.

A precision and recall analysis was done on the results to confirm the accuracy of the model in finding the two clusters that represent the Democratic and Republican parties. Precision is defined as the probability of belong to a party given the individual is classified under its corresponding cluster (cluster 1 corresponds to the Democratic party, and cluster 2 the Republican Party). Recall is defined as the probability of classifying under a cluster given the individual belongs to its corresponding party.

| | **Democrat/Cluster 1** | **Republican/Cluster 2** |
|---|---|---|
| Precision | 76% | 73% |
| Recall | 89% | 52% |

**Table 2: Year 2012 – Precision and Recall Percentages for GMM Model.** This table includes resulting precision and recall for the two clusters found by the Gaussian Mixture Model given the self-identified party affiliation

A possible explanation for the discrepancy of Republican Party position and the right-wing cluster mean is that since the right-wing cluster is small in size, the Republican Party needs to attract centrist voters to be successful at elections. Thus it takes on a mixture of center-right and right-wing positions.

We have explored more two-issues combinations, and they are reported in Appendix B.

**Year 2012, Eleven Issues**

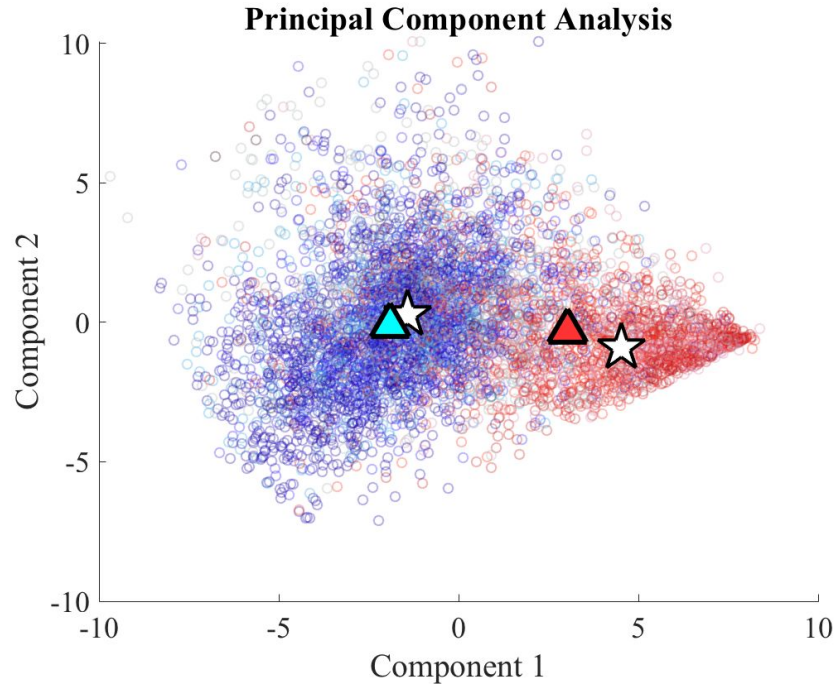

**Figure 2: Year 2012 – Eleven Issues, Two Clusters.** The stars are where the cluster means were found. The two triangles represent the political party means; red for republican and blue for democrat. These were found by taking the average of the opinions of those that self-identified as democrat and as republican. Principal Component Analysis was then used to plot the points onto two dimensions.

The two political issue analysis was then applied to higher dimensions in order to model a more complex opinion space. The multidimensional opinion space was graphed on two dimensions found through Principal Component Analysis, which finds two axes with the most variation in data. The data is then projected onto these two axes. The results from this analysis are shown in Figure 2. Even with 11 political issues, the positions of the two clusters are consistent with those found with two political issues. One cluster is found very close to the Democratic Party mean, and another cluster is found farther right than the Republican Party mean. In the 11-issue analysis, we found a large and a small cluster similar to the two-issue analysis. The large cluster is close to the Democratic party position and contains 76% of the individuals. The small cluster is more extremist than both the Democratic and Republican party position and contains 24% of the population (AIC = $2.68 \times 10^5$).

**Time-Series 2012-1990 - Three Issues**

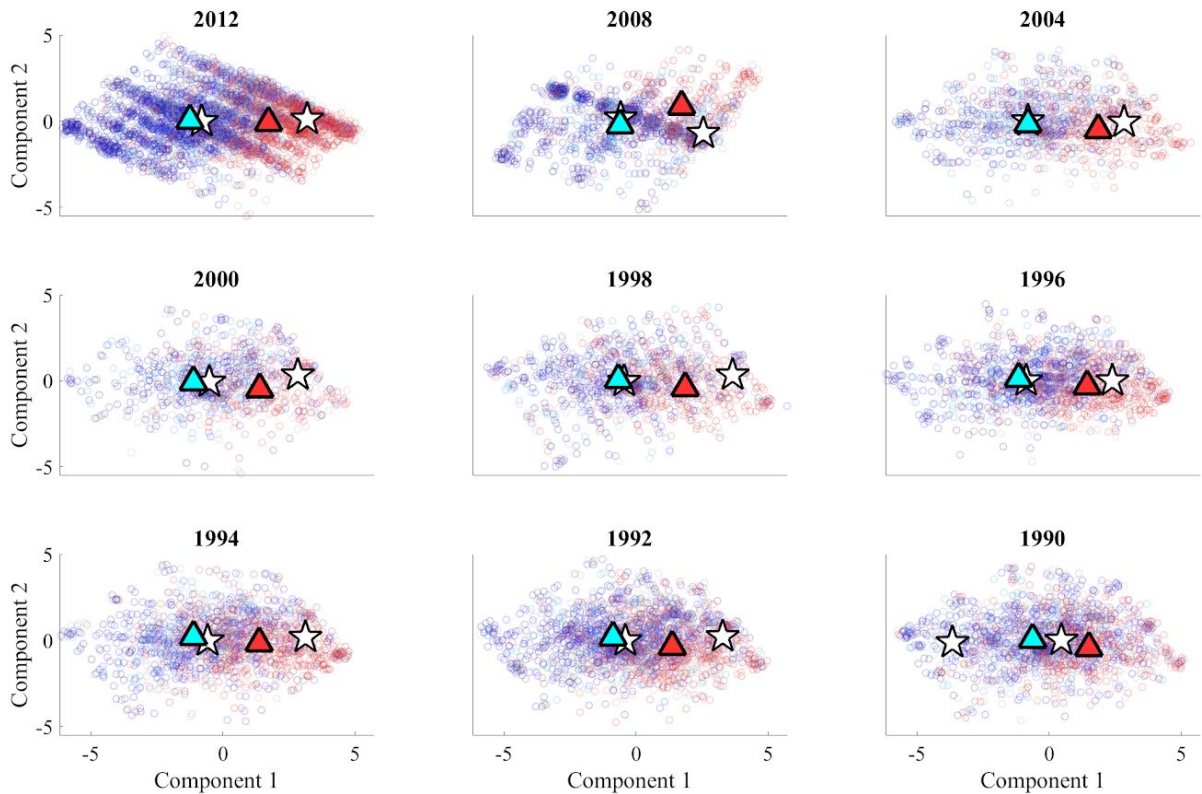

**Figure 3: 2012-1990 – 3 Issues, 2 Clusters.** The stars are where the cluster means were found. The two triangles represent the political party means; red for republican and blue for democrat. These were found by taking the average of the opinions of those that self-identified as democrat and as republican. Principal Component Analysis was then used to plot the points onto two dimensions.

The two-cluster analysis was then repeated for nine years between 1990 and 2012 using three political issues occurring in surveys of all years. See Appendix A for details on the political issues. The results shown in Figure 3 are consistent with those in Figure 1 and 2. Except in 2008 and 1990. The deviation in 2008 may be explained by the lack of data collected in that year. The year 1990 is an exception --- one cluster is found way left of the Democratic Party mean and another in the middle of the two party means.

Two visualizations are shown in Figure 4 and Figure 5, the distance between the cluster means and party means, and the distance from the center. The distance between the cluster means is significantly greater than the distance between the party means. This suggests that the Gaussian Mixture Model found that the public opinion is better represented when the two clusters are farther apart than where the party means currently lie. It is also suggestive that the distance between the party means is increasing over time. This indicates that the view of the Republican Party may be migrating to the right, where the small cluster was found. Figure 5 displays the distance of each cluster and party mean from the center. This graph supports the claim that the Republican Party has consistently taken a more moderate stance than that of some of its supporters.

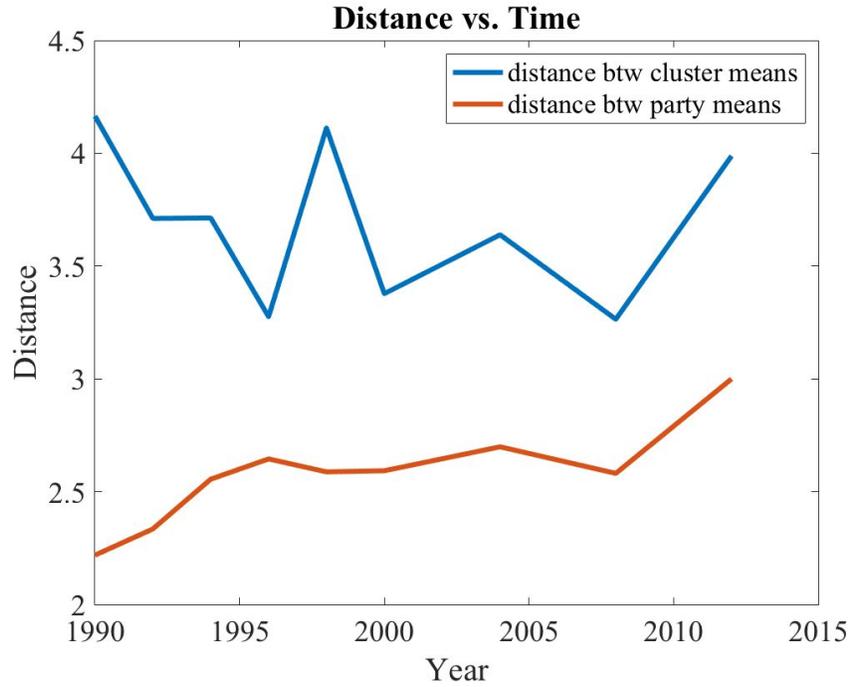

**Figure 4: 2012-1990 – 3 Issues, 2 Clusters. Comparison of Distance between Cluster Means and Party Means.** The blue line represents the distance between the cluster means and the red line represents the distance between the party means.

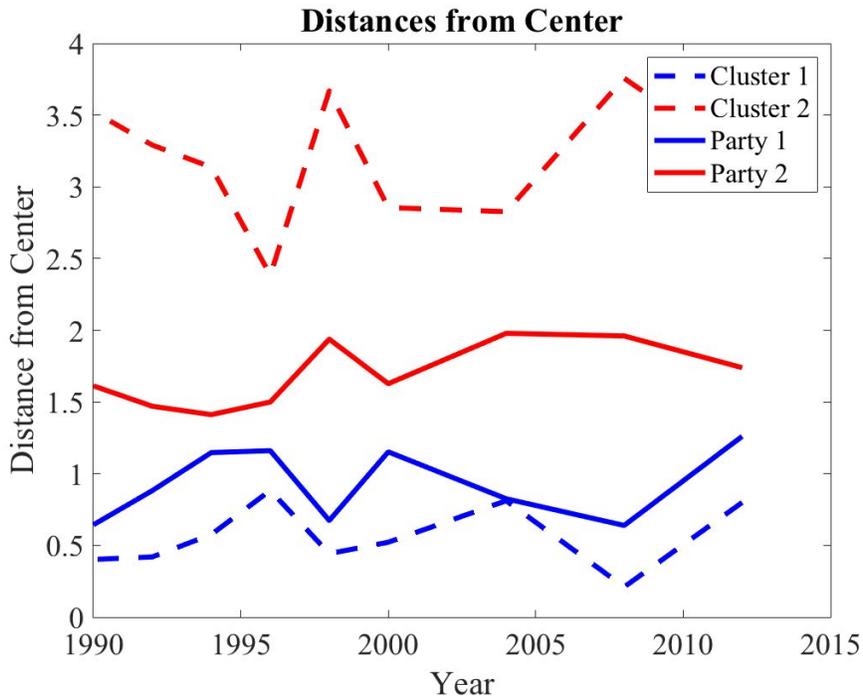

**Figure 5: 2012-1990 – 3 Issues, 2 Clusters. Comparison of Distance from Center.** The dashed lines represent the distance from a cluster mean to the center. The solid lines represent the distance from one party mean to the center (blue for democrat, red for republican).

## More than Two Clusters

This section focuses on relaxing the two-cluster constraint and discovering the optimal number of clusters that represent the individuals in the dataset. The analysis was conducted on the 2012 – two issue dataset and repeated three times for validation of the results. We evaluate the models by the AIC value.
We added a constraint to the model that the variance values (diagonal elements of $\Sigma_i$) should be greater than 1. The reason is that the data on the 7-point discrete scale can often generate "single position clusters" (with near zero variance) as the number of clusters are allowed to be large. However, we think that a reasonable cluster representing an ideology group should be broad and not depend on whether an individual answered "3" or "4" on a particular question, which is both unrealistic and not robust to uncertainty in people reporting to surveys. The results indicate that 3 clusters are a better representation of the data than 2 and 4 clusters. This is shown through the local minimum of the AIC value, shown in Table 3. The resulting contour graph is shown in Figure 6. The result is robust, as the covariance values found for these 3 clusters are not hovering very close to the covariance constraint. The locations of the 3 clusters lie almost along the diagonal of the opinion space, with a left group, a centrist group near the current Democratic Party, and a right group that is further right to the current Republican Party. Table 4 displays the center positions and variances of the clusters found, as well as the size and party composition of the clusters.

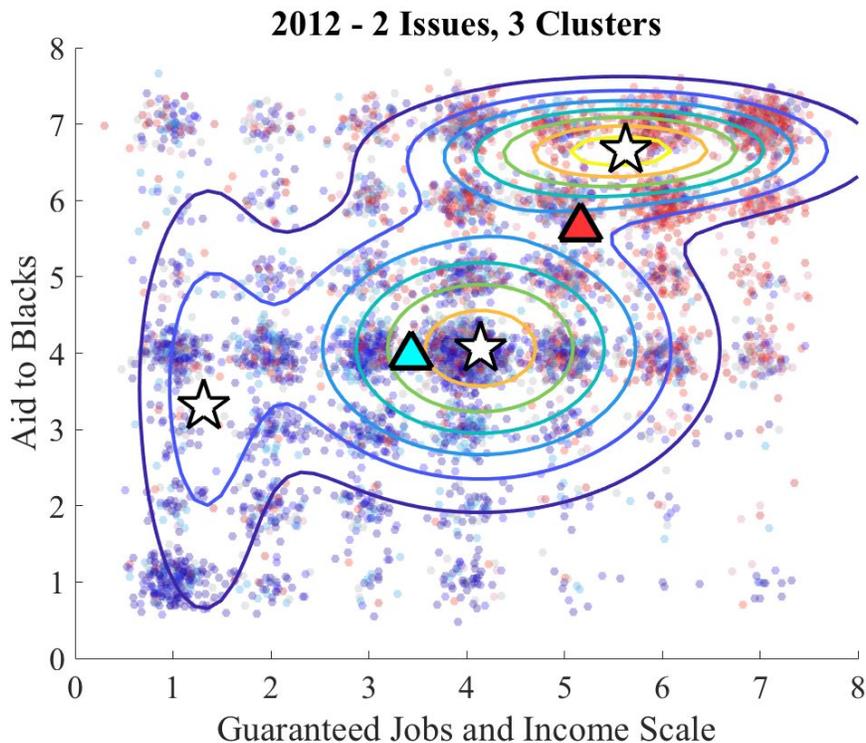

**Figure 6: Year 2012 – Two Issues, Three Clusters.** The stars represent the cluster means. The two triangles represent the political party means; red for republican and blue for democrat. These were found by taking the average of the opinions of those that self-identified as democrat and as republican.

| AIC |
|---|

| | |
|---|---|
| 2 clusters | $4.87 \times 10^4$ |
| 3 clusters | $4.68 \times 10^4$ |
| 4 clusters | $4.80 \times 10^4$ |

**Table 3: Year 2012 – Two Issues.** AIC value comparisons between 2, 3, and 4 clusters.

| | **Cluster 1** | **Cluster2** | **Cluster3** |
|---|---|---|---|
| Cluster center position | (1.5, 3.2) | (3.7, 4.2) | (5.5, 6.2) |
| Variance | (1.7, 2.0) | (1.9, 1.3) | (1.4, 1.0) |
| Contains portion of data | 13% | 58% | 29% |
| Democrat | 63% | 46% | 17% |
| Independent | 29% | 37% | 39% |
| Republican | 8% | 17% | 44% |

**Table 4: Year 2012 – Two Issues, Three Clusters.** The cluster center position, variance, size and partisan composition of the clusters.

## Discussions

Motivated by the two-party political landscape in the US, we first performed clustering of public opinion data into two clusters to see how well do the two parties represent the public opinion. In the public opinion data, we find a large centrist cluster and a small right-wing cluster in the public opinion. The Democratic Party position lies close to the centrist cluster, and the Republican Party position lies between the centrist cluster and the right-wing cluster.

After removing the two-cluster limitation, we find that three clusters are the best representation of the data – a centrist clusters, a left clusters, and a right cluster. We also used a number of modifications of the model (see Appendix C) for robustness, and found that the more than two parties always represent the population better than two parties, even accounting for additional parameters.

In this research, we considered other unsupervised clustering techniques such as k-means. However, we did not use k-means because it favors solutions with clusters of similar sizes. We would like to leave open the possibility of finding co-existing large and small clusters (which can represent large and small interest groups, as the ones we found the two-cluster analyses). Thus we chose the Gaussian Mixture Model, which does not have this bias.

In the three-cluster analysis, we find two of the three clusters are dominantly Democrats, and the Republican majority cluster in the three-cluster analysis (reported in Table 4) is of similar position and size as that in the two-cluster analysis (as reported in Table 1). This is suggestive that the two Democrat-majority clusters found in the three-cluster analysis may be segmentations of the large cluster found in the two-cluster analysis.

This research suggests that there are better representations of the political ideology of the public than the current two-party landscape. When constrained two-clusters, we found two size imbalanced clusters. The Republican party is representing both some centrist and most ring-wing individuals. We find that three parties give the best representation of the public opinion data.

## Limitations and Future Work

Some challenges still remain for a robust estimate of the optimal number of clusters, given the low-resolution of the data in discrete categories. Here, we arrive at the estimate of three clusters, under the assumption that such clusters representing political interests should be reasonably broad, with variance spanning more than one point on the seven-point scale. A challenge still remains is that when this analysis was done for a greater number of clusters, the covariance values began to straddle the constraint. This indicates that the clusters found are being limited by the constraint. Subsequently, more research needs to be done in determining the value to set the constraint that corresponds to a political platform. The analysis for the optimal number of clusters was also limited to data on two political issues. However, the public opinion space is not based on only two issues. In the future, the two dimensional analysis should be applied to higher dimensions to make sure that the clusters found are not dependent on one political issue. Another limitation of this analysis lies in fact that each political issue has equal weight in determining the clusters. In truth, there are some political issues that voters prioritize and deem more important than others. These issues should have more weight in determining where the clusters form. Another main limitations lie in the discreteness of the data. In the future, methods of dealing with discrete datasets in statistical analysis should be researched.

Recent political science research shows consistent discrepancies between a constituency ideology and that of their representative's perception [7]. We hope that more quantitative research on public ideology will help bridge this gap. We understand there are many systematic barriers in the US for additional parties. Some examples are the plurality election system, which is known to favor a two-party system [8]; minimum requirements for receiving presidential campaign funding (over 5% national polling) and being in the presidential debate (over 15%). In our analysis, we find that the public opinions are in nature more than two clustered. We also encourage further research on how a system that prefers a two-party system affect political representation in a multi-cluster ideology landscape.

## Acknowledgement

The authors would like to thank Daniel Abrams, Adilson Motter, and Doug Downey for guidance and helpful discussions. The authors would like to acknowledge support from Weinberg Research Grant from Northwestern University that supported this research.## Author Contributions

L.L. wrote the clustering program, performed analysis, and generated all the results and figures that appeared in the main text, appendix A & C. L.L. also wrote the initial draft of the manuscript and contributed to edits of the final manuscript. S.Z. performed initial exploratory analysis and generated Appendix B. V.C.Y. conceived the research question, designed the research method and edited the final manuscript. L.L. and V.C.Y discussed the results.

# Appendix A: Political Issues

## 2012- 2 Political Issues

| Political Issue | **Guaranteed Jobs and Income Scale** | **Aid to Black Scale** |
|---|---|---|
| Question Asked in Survey | "Some people feel that the government in Washington should see to it that every person has a job and a good standard of living Others think the government should just let each person get ahead on his/their own.<br><br>Where would you place yourself on this scale, or haven't you thought much about this?" | "Some people feel that the government in Washington should make every (prior to 1996 only: possible) effort to improve the social and economic position of blacks. Others feel that the government should not make any special effort to help blacks because they should help themselves.<br><br>Where would you place yourself on this scale, or haven't you thought much about it?" |
| Scale | 0 = NA<br>1 = Gov. should see to job and good standard of living<br>7 = Gov. should let each person get ahead on his own<br>9 = DK | 0 = NA<br>1 = Gov. should help minority groups<br>7 = Minority groups should help themselves<br>9 = DK |

## 2012 – 11 Political Issues

| Political Issue | **Government Health Insurance Scale** | **Guaranteed Jobs and Income Scale** | **Aid to Black Scale** |
|---|---|---|---|
| Survey Question | "Some (1994-later: people) feel there should be a government insurance plan which would cover all medical and hospital expenses for everyone. Others feel that (1994-1996: all) medical expenses should be paid by individuals, and through private insurance plans like Blue Cross (1984-1994: or [1996: some] other company paid plans).<br><br>Where would you place yourself on this scale, or haven't you thought much about this?" | "Some people feel that the government in Washington should see to it that every person has a job and a good standard of living Others think the government should just let each person get ahead on his/their own.<br><br>Where would you place yourself on this scale, or haven't you thought much about this?" | "Some people feel that the government in Washington should make every (prior to 1996 only: possible) effort to improve the social and economic position of blacks. Others feel that the government should not make any special effort to help blacks because they should help themselves.<br><br>Where would you place yourself on this scale, or haven't you thought much about it?" |
| Scale | 0 = NA<br>1 = Gov. insurance plan<br>7 = Private insurance plan<br>9 = DK | 0 = NA<br>1 = Gov. should see to job and good standard of living<br>7 = Gov. should let each person get ahead on his own<br>9 = DK | 0 = NA<br>1 = Gov. should help minority groups<br>7 = Minority groups should help themselves<br>9 = DK |

| Political Issue | **Government Service-Spending Scale** | **Defense Spending Scale** | **President on Defense Spending Scale** |
|---|---|---|---|
| Survey Question | "Some people think the government should provide fewer services, even in areas such as health and education, in order to reduce spending. Other people feel that it is important for the government to provide many more services even if it means an increase in spending. Where would you place yourself on this scale, or haven't you thought much about this?" | "Some people believe that we should spend much less money for defense. Others feel that defense spending should be greatly increased. Where would you place yourself on this scale or haven't you thought much about this?" | "Some people believe that the President should spend much less money for defense. Others feel that defense spending should be greatly increased. What should the President do in regards to Defense Spending or haven't you thought much about this?" |
| Scale | 0 = NA<br>1 = Gov. should provide many fewer services: reduce<br>7 = Gov. should provide many more services: increase<br>9 = DK | 0 = NA<br>1 = Greatly decrease defense spending<br>7 = Greatly increase defense spending]<br>9 = DK | 0 = DK<br>1 = Greatly decrease defense spending<br>7 = Greatly increase defense spending<br>9 = NA |
| Political Issue | **President on Aid to Blacks Scale** | **President on Government Health Insurance Scale** | **President on Government Spending/Services Scale** |
| | "Some people feel that the President should make every (prior to 1996 only: possible) effort to improve the social and economic position of blacks. Others feel that the President should not make any special effort to help blacks because they should help themselves. Where should the President do on this scale, or haven't you thought much about it?" | "Some (1994-later: people) feel there should be a government insurance plan which would cover all medical and hospital expenses for everyone. Others feel that (1994-1996: all) medical expenses should be paid by individuals, and through private insurance plans like Blue Cross (1984-1994: or [1996: some] other company paid plans). What should the President do on this scale, or haven't you thought much about this?" | "Some people think the government should provide fewer services, even in areas such as health and education, in order to reduce spending. Other people feel that it is important for the government to provide many more services even if it means an increase in spending. Where should the President do on this scale, or haven't you thought much about this?" |
| Scale | 0 = DK<br>1 = Gov. should help minority blacks/minorities<br>7 = Blacks/minorities should help themselves<br>9 = DK | 0 = DK<br>1 = Government insurance plan<br>7 = Private insurance plan<br>9 = NA | 0 = DK<br>1 = Gov. should provide many fewer services: reduce<br>7 = Gov. should provide many more services: increase<br>9 = NA |
| Political Issue | **President on Guaranteed Jobs and Living Scale** | **President on Liberal Scale** | |
| Survey Question | "Some people feel that the government in Washington should see to it that every person has a job and a good standard of living Others think the government should just let each person get ahead on his/their own.<br><br>What should the President do on this scale, or haven't you thought much about this?" | "Where would you place the President on a Liberal-Conservative scale, or haven't you thought much about it?" | |
| Scale | 0 = DK<br>1 = Gov. should see to job and good standard of living<br>7 = Gov. should let each person get ahead on his own<br>9 = NA | 0 = DK<br>1 = Extremely liberal<br>2 = Liberal<br>3 = Slightly liberal<br>4 = Moderate<br>5 = Slightly conservative<br>6 = Conservative<br>7 = Extremely conservative<br>9 = NA | |

**2012-1990 – 3 Political Issues**

| Political Issue | **Guaranteed Jobs and Income Scale** | **Aid to Black Scale** | **Government Service-Spending Scale** |
|---|---|---|---|
| Survey Question | "Some people feel that the government in Washington should see to it that every person has a job and a good standard of living Others think the government should just let each person get ahead on his/their own.<br><br>Where would you place yourself on this scale, or haven't you thought much about this?" | "Some people feel that the government in Washington should make every (prior to 1996 only: possible) effort to improve the social and economic position of blacks. Others feel that the government should not make any special effort to help blacks because they should help themselves.<br><br>Where would you place yourself on this scale, or haven't you thought much about it?" | "Some people think the government should provide fewer services, even in areas such as health and education, in order to reduce spending. Other people feel that it is important for the government to provide many more services even if it means an increase in spending.<br><br>Where would you place yourself on this scale, or haven't you thought much about this?" |
| Scale | 0 = NA<br>1 = Gov. should see to job and good standard of living<br>7 = Gov. should let each person get ahead on his own<br>9 = DK | 0 = NA<br>1 = Gov. should help minority groups<br>7 = Minority groups should help themselves<br>9 = DK | 0 = NA<br>1 = Gov. should provide many fewer services: reduce<br>7 = Gov. should provide many more services: increase<br>9 = DK |

# Appendix B: Additional results for two-issue combinations

**2012- 2 Political Issues**

Combination 1:
Issue 1: Government Health Insurance Scale
Issue 2: Guaranteed Jobs and Income Scale
Correlation coefficient: 0.4872

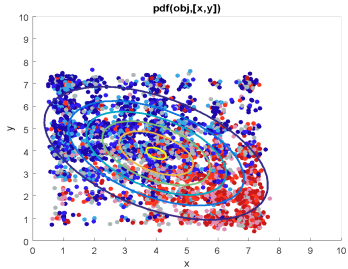 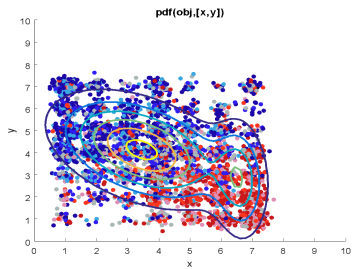 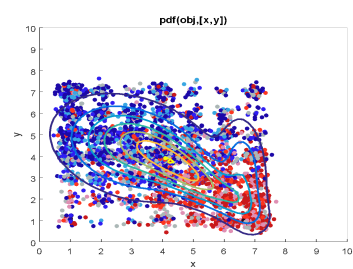

Figure: 1-Cluster        Figure: 2-Cluster        Figure: 3-Cluster

| 1-cluster | | 2-cluster | | 3-cluster | |
| --- | --- | --- | --- | --- | --- |
| AIC | BIC | AIC | BIC | AIC | BIC |
| 4.64E+04 | 4.65E+04 | 4.56E+04 | 4.56E+04 | 4.50E+04 | 4.51E+04 |

Combination 2:
Issue 1: Guaranteed Jobs and Income Scale
Issue 2: Aid to Blacks Scale
Correlation coefficient: 0.4861

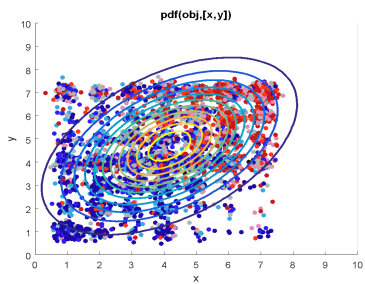 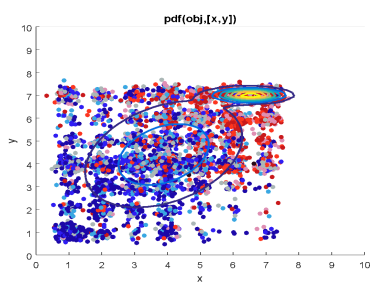

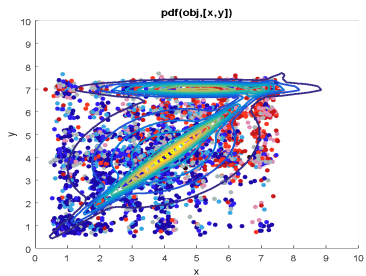

Figure: 1-Cluster        Figure: 2-Cluster        Figure: 3-Cluster

| 1-cluster | | 2-cluster | | 3-cluster | |
| --- | --- | --- | --- | --- | --- |
| AIC | BIC | AIC | BIC | AIC | BIC |
| 4.53E+04 | 4.54E+04 | 4.39E+04 | 4.40E+04 | 4.27E+04 | 4.28E+04 |

Combination 3:
Issue 1: Government Health Insurance Scale
Issue 2: Defense Spending Scale
Correlation coefficient: 0.2923

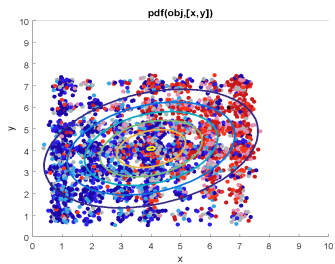
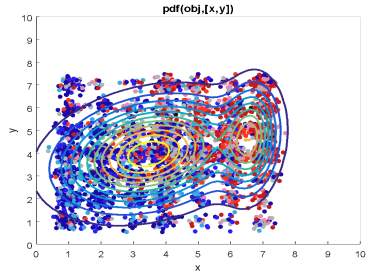

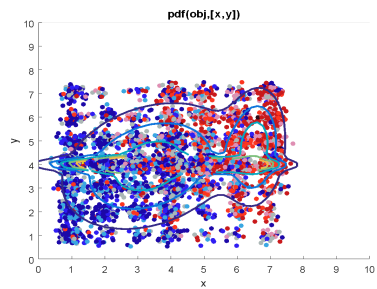

| Figure: 1-Cluster | | Figure: 2-Cluster | | Figure: 3-Cluster | |

| 1-cluster | | 2-cluster | | 3-cluster | |
| --- | --- | --- | --- | --- | --- |
| AIC | BIC | AIC | BIC | AIC | BIC |
| 4.51E+04 | 4.51E+04 | 4.45E+04 | 4.46E+04 | 4.29E+04 | 4.30E+04 |

Combination 4:
Issue 1: Government Health Insurance Scale
Issue 2: Government Services-Spending Scale
Correlation coefficient: -0.4843

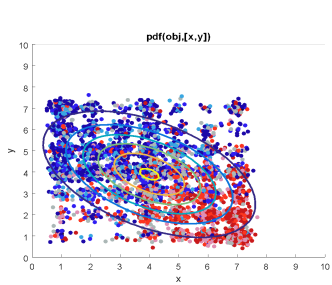
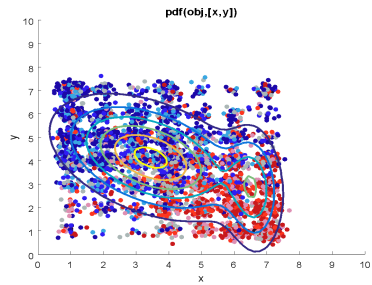
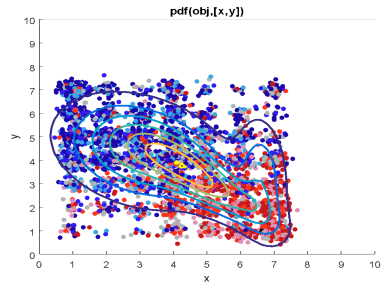

| Figure: 1-Cluster | Figure: 2-Cluster | Figure: 3-Cluster |

| 1-cluster | | 2-cluster | | 3-cluster | |
| --- | --- | --- | --- | --- | --- |
| AIC | BIC | AIC | BIC | AIC | BIC |
| 4.49E+04 | 4.50E+04 | 4.42E+04 | 4.43E+04 | 4.39E+04 | 4.40E+04 |

It is observed that when two combinations have similar correlation coefficients, the graphs are similar. Even though sometimes the correlation coefficients are different, the difference does not influence the results of AIC and BIC. The only thing changed is the slope of contour, which is the same as the correlation coefficients. Therefore, the effect of correlation can be excluded.

## Appendix C: Additional discussion on more than two cluster analysis

Here we performed two types of analysis before constraining the covariance value to be larger than one. These analysis, though not directly related to the result, illustrates some difficulties in working with the discrete data and motivates us to add the constraint on the covariance.

**First analysis – direct comparison**
The first analysis directly compares the AIC calculated from the likelihood generated by the Gaussian Mixture Model. The results indicate that 6 clusters is a good representation of the data, shown in Figure C1. However, we notice a potential issue with this direct comparison of AIC values. As illustrated in Table C1, the clustering result finds near zero covariance values along some dimensions. We think this may be an artifact from the seven-point scale discretization of our data. Near zero covariance suggests that individuals who only differ by one point on an issue (such as reporting three vs. four) are clustered in different clusters. This seems both unrealistic and not robust to reporting uncertainty in surveys (as an individual can sometimes report three and sometimes four to the same question). As the number of clusters increase, we see more and more of these "single position clusters". We would like to make sure that we are not overfitting the data thus we perform the second analysis.

**Second analysis – cross validation**
In check for overfitting, we used a 5-fold cross validation. Four-fifth of the data was used for training; the Gaussian Mixture Model was implemented to find a set of parameters that maximized the likelihood value. These parameters were then used to calculate the likelihood of the testing data, the leftover 1/5 of the data. This procedure was iterated for up to 20 clusters. Figure C2 shows the AIC values calculated for each cluster.

We find that as we increase the number of clusters allowed in the Gaussian Mixture model, the AIC decreases, suggesting better fit for data even when penalizing additional parameters. If overfitting does occur, we would expect to see an increase in AIC at the number of clusters which over fitting happens. I think here, we are looking at an outcome of clustering low resolution, discrete data. As the number of clusters get large, more integer points on the seven-by-seven grid tend to become its own clusters. We do not think this type of clustering is a useful representation of the data, however, it would not be reflected by AIC values.

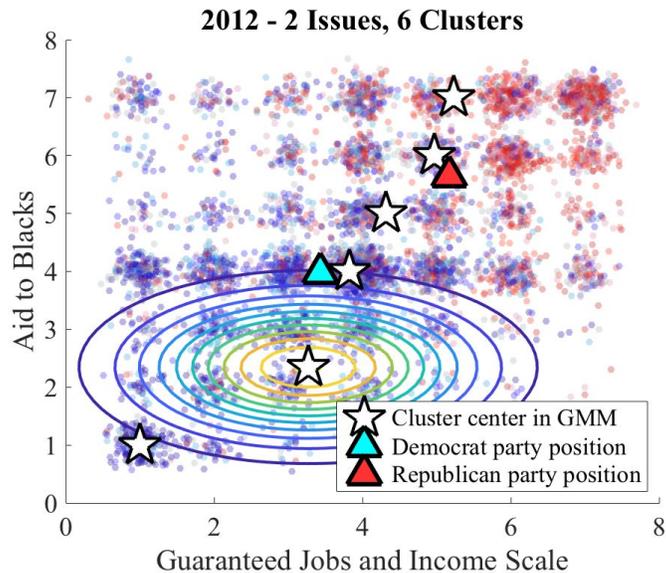

**Figure C1: 2012 – Two Issues, Six Clusters.** The stars are where the cluster means were found. The two triangles represent the political party means; red for republican and blue for democrat. These were found by taking the average of the opinions of those that self-identified as democrat and as republican.

| Iterations | 40 | |
|---|---|---|
| Log-likelihood | 4227.26 | |
| Mean | 3.2690 | 2.3457 |
| | 5.2306 | 7.0000 |
| | 1.0000 | 1.0000 |

|  | 4.9685   6.0000 |
|---|---|
|  | 4.3149   5.0000 |
|  | 3.8281   4.0000 |
| Covariance | 1.9810   0.5765 |
|  | 3.3060   0.0000 |
|  | 1.0e-05 *   1.0000 1.0000 |
|  | 2.0842   0.0000 |
|  | 2.0784   0.0000 |
|  | 2.1873   0.0000 |
| AIC | -8.40E+03 |

**Table C1: 2012 – 2 Issues, 2 Clusters.** This table includes the number of iterations it took in the Gaussian Mixture Model to get the results. It shows the Log-likelihood value that was maximized, the resulting mean and diagonal covariance values, and the Akaike Information Criterion.

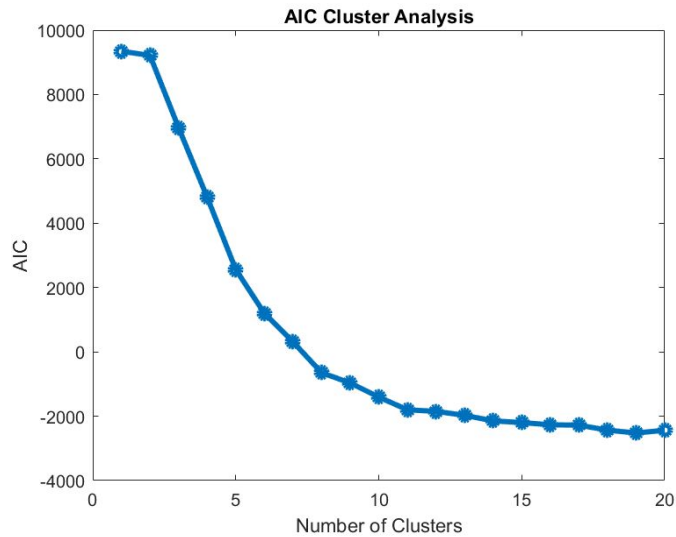

**Figure C2: 2012 – 2 Issues. Comparison of AIC using Cross Validation.** The figure displays the AIC values calculated from the likelihood generated through a 5-fold cross validation up to 20 clusters.